\documentclass[usenatbib]{mn2e}
\usepackage{amssymb}
\usepackage{psfig}
\usepackage{amsmath}
\footnotesize

\newcommand\nuddd{\ifmmode\stackrel{\bf \,...}{\textstyle \nu}\else$\stackrel{\,...}{\textstyle \nu}$\fi}
\def\lsim{~\rlap{$<$}{\lower 1.0ex\hbox{$\sim$}}}
\def\gsim{~\rlap{$>$}{\lower 1.0ex\hbox{$\sim$}}}

\title{ The environment and star formation of HII region Sh2-163: a multi-wavelength study}

\author[Naiping Yu, Jun-Jie Wang and Nan Li]{Naiping Yu$^{1,2}$\thanks{E-mail: yunaiping09@mails.gucas.ac.cn},
Jun-Jie Wang$^{1,2}$ and Nan Li$^{1,2}$\footnotemark[1]\thanks{}\\
$^{1}$National Astronomical Observatories, Chinese Academy of
Sciences, Beijing 100012, China\\
$^{2}$NAOC-TU Joint Center for Astrophysics, Lhasa 850000, China}

\begin{document}
\maketitle
\pagestyle{plain}

\begin{abstract}
In order to investigate the environment of HII region Sh2-163 and
search for evidence of triggered star formations in this region, we
performed a multi-wavelength study of this HII region. Most of our data were
taken from large-scale surveys: 2MASS,
CGPS, MSX and SCUBA. We also made CO molecular line
observations, using the 13.7m telescope. The ionized region of
Sh2-163 is detected both by the optical and radio continuum
observations. Sh2-163 is partially bordered by an arc-like
photodissociation region (PDR), which is coincident with the
strongest optical and radio emissions, indicating
interactions between the HII region and the surrounding interstellar
medium (ISM). Two molecular clouds were discovered on the border of
PDR. The morphology of these two clouds suggests they are compressed by
the expansion of Sh2-163. In cloud A, we found two molecular clumps.
And it seems star formation in clump A2 is much more active than in clump A1.
In cloud B, we found new outflow activities and massive star(s)
are forming inside. Using 2MASS photometry, we tried to search for
embedded young stellar object (YSO) candidates in this region. The
very good relations between CO emission, infrared shell and YSOs
suggest that it is probably a triggered star formation region by the
expansion of Sh2-163. We also found the most likely massive protostar
related to IRAS 23314+6033.

\end{abstract}


\section{Introduction}
A lot of research has been done in astrophysics to understand the
formation of massive stars and the feedback to their surrounding
ISM (e.g. Zinnecker et al. 2007; Deharveng et al. 2010, and references therein).
However, many questions are still unclear. Multi-wavelength observations
are essential to deeply understand how the triggered star formation processes
impact on the massive star formation process.
Young massive stars tend to form in clusters or
groups. The formation of massive stars has an immense impact on
their environment through ionizing radiation, heating of dust and
expansions of their HII regions. These processes may trigger next
generation of star formation by compressing neighboring molecular
clouds to the point of gravitational instability. Massive stars also
have powerful winds which sweep up the surrounding gas, creating
interstellar bubbles (e.g. Weaver et al. 1977; Churchwell et al. 2006). In the case of OB associations,
the released intense ultraviolet radiation may ionize the surrounding ISM
within tens of parsecs. A number of observations demonstrate HII regions
can strongly affect star formation nearby. Sugitani et al. (1989)
showed the ratios of luminosity of the protostar to core mass in
bright-rimmed clouds are much higher than those in dark globules.
Dobashi et al. (2001) also showed that protostars associated with
HII regions are more luminous than those in molecular clouds away
from them, indicating HII regions favour massive stars or cluster
formation in neighboring molecular clouds. Moreover, in the HII
region of W5, Karr and Martin (2003) found the number of
star-formation events per unit CO covering area within the influence
zone is 4.8 higher than outside. However, the role of expanding HII
regions in triggering star formation is still poorly understood. For
example, Dale et al. (2007a,b) argue that the main effect of an
expanding HII region may simply be to expose stars that would have
formed anyway.

Several mechanisms by which massive stars can affect the
subsequent star formation in an HII region have been proposed. Two
of the most studied processes are known as ``radiatively driven implosion''
(RDI) (e.g. Lefloch \& Lazareff 1994; Miao et al.2006; Miao et al.
2009 ) and ``collect and collapse'' (C\&C) (e.g. Elmegreen \& Lada
1977; Osterbrock 1989). According to the model of ``RDI'', the expanding
ionization fronts caused by the UV radiations impact into
pre-existing molecular clouds, leading to the formation of a
cometary globule, where new stars may finally be born (Larosa 1983). The ``C\&C'' model invokes the
standard picture of a slow moving D type ionization front with
associated shock front that precedes the ionization front
(Osterbrock 1989). Dense gas may pile up between the two fronts. On
a long time the compressed shocked layer becomes gravitationally
unstable and then star formation will take place inside. Observational
evidence of both processes has been proposed in a number of HII regions
(e.g. Deharveng \& Zavagno 2008; Cichowolski et al. 2009; Paron et
al. 2011). In this paper, we made a multi-wavelength study of Sh2-163 to
find out whether second-generation clusters are forming around. We also
discussed the physical mechanisms which may trigger star formation in this region.

Sh2-163 is an optically visible HII region centered on R.A. (2000) =
23h32m57.9s and Dec. (2000) =
60$^\circ$48$^\prime$01$^\prime$$^\prime$ with a mean diameter of
about 10$^\prime$ (Sharpless 1959). The distance to Sh2-163 has been
estimated by several authors, using different methods. CO observations
by Blitz et al. (1982) shows it has a velocity of -44.9 $\pm$ 3.8 km/s
(the local standard of rest velocity), corresponding to
a kinematic distance of 2.3 $\pm$ 0.7 kpc (Brand \& Blitz 1993). A
spectrophotometric distance of 2.7 $\pm$ 0.9 kpc was derived by
Georgelin (1975). According to Russeil et al. (2007), Sh2-163
belongs to the complex 114.0 - 0.7, which is composed of Sh2-163,
Sh2-164, and Sh2-166. All of the three HII regions are located on the
Norma-Cygnus arm. Based on spectroscopic
and UBV photometric observations, Russeil et al. (2007) found Sh2-163 is ionized by an O9V star (R.A. (2000) =
23h33m36.9s, Dec. (2000) =
60$^\circ$45$^\prime$06.8$^\prime$$^\prime$) and an O8V star (R.A.
(2000) = 23h33m32.7s, Dec. (2000) =
60$^\circ$47$^\prime$32.1$^\prime$$^\prime$). And the two ionizing stars have a mean distance of 3.3 $\pm$ 0.3 kpc.
Thus, the distance of Sh2-163 is in the range of 2.3 to 3.3 kpc. We take a
distance of 2.8 $\pm$ 0.5 kpc in the following discussions.

\begin{figure}
\centerline{\psfig{file=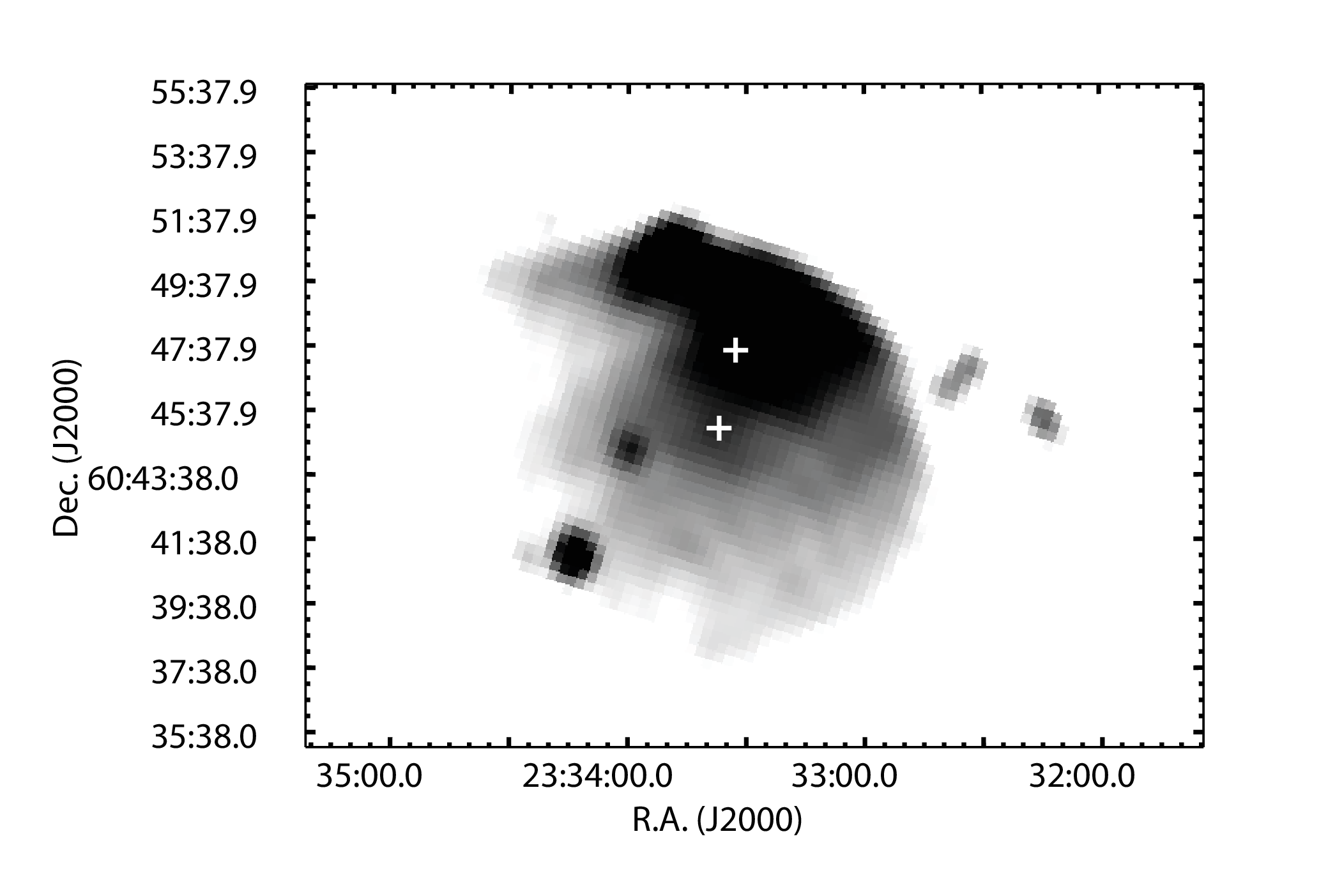,width=2.6in,height=1.8in}}
\centerline{\psfig{file=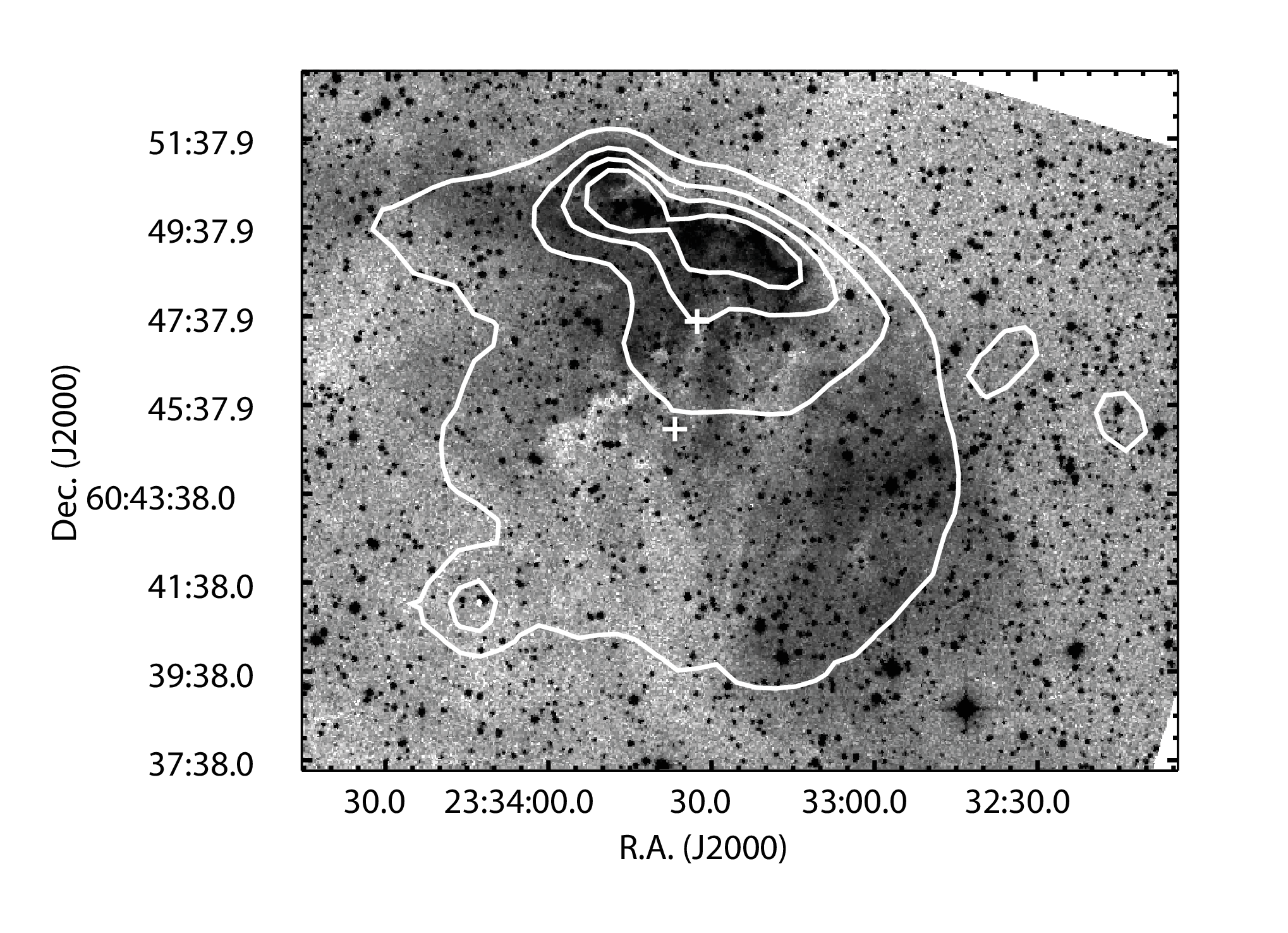,width=2.6in,height=1.8in}}
\centerline{\psfig{file=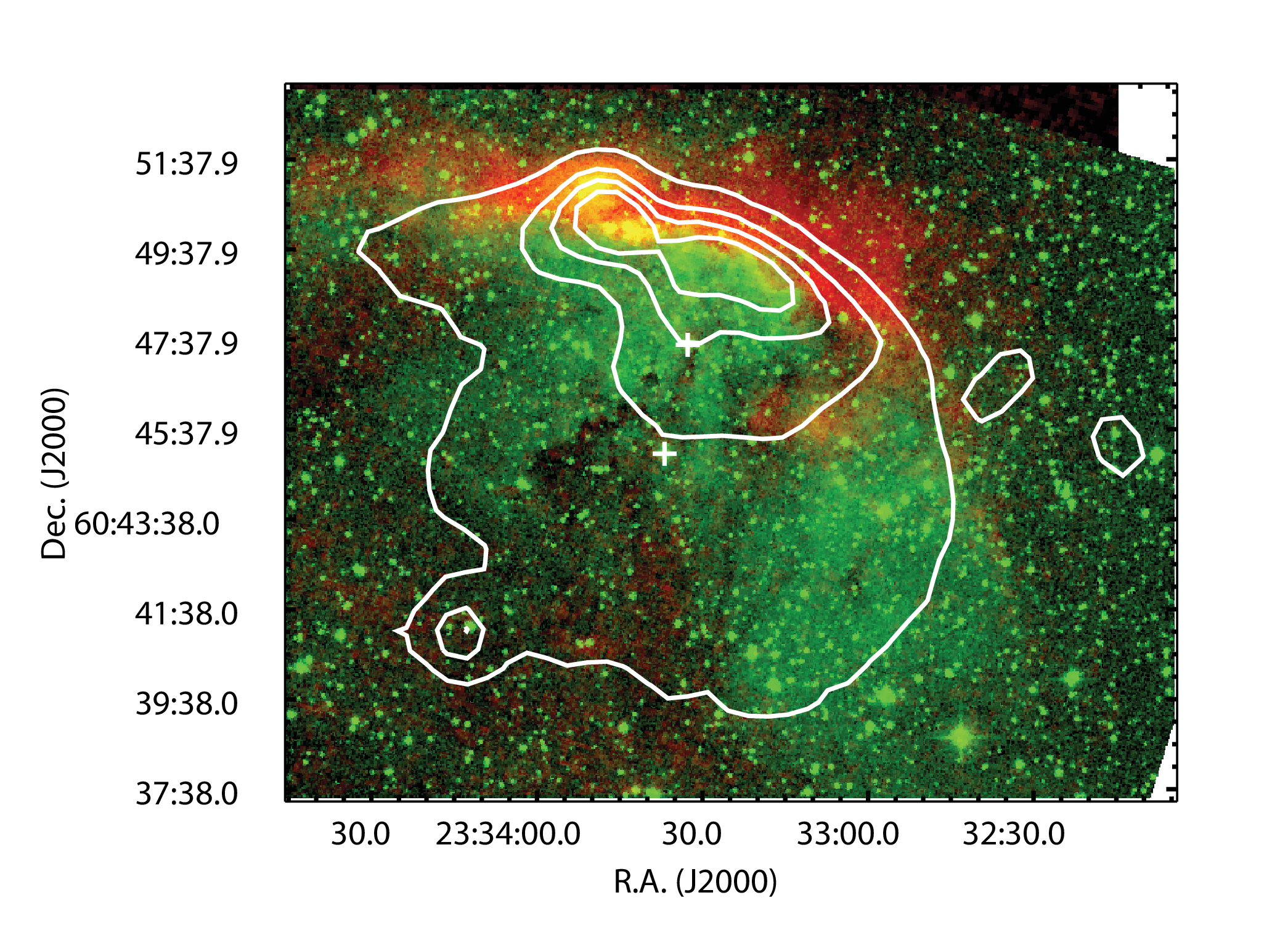,width=2.6in,height=1.8in}}
\caption{Top:1420 MHz image of Sh2-163. The two white pluses mark the
ionizing stars found by Russeil et al. (2007). Middle: DSS-R image
(grey-scale) with the 1420 MHz radio continuum contours. Contour
levels are 8, 11, 14 and 17 K. Bottom: A two-color image of Sh2-163:
DSS-R image (green) and MSX band A image (red). The contours are the
same as those in the middle panel. }
\end{figure}

\section{Data Sets and Observations}
The Canadian Galactic Plane Survey (CGPS) is a project combining
radio, millimeter, and infrared surveys of the Galactic plane. The
radio surveys were carried out at the Dominion Radio Astrophysical
Observatory (DRAO), covering the region 74. $^{\circ}$2 $<$ $\ell$
$<$ 147. $^{\circ}$3, with Galactic latitude extent -3. $^{\circ}$6 $<$
$\mathit{b}$ $<$ +5.$^{\circ}$6 at 1420 MHz (Taylor et al. 2003),
resolving features as small as 1 arcminute. In order to match the
DRAO images, the CGPS data base also comprises other data sets, such
as the Five College Radio Astronomical Observatory (FCRAO) CO(1-0)
Survey of the Outer Galaxy (Heyer et al. 1998).

Mid-IR data were taken from the Midcourse Space Experiment (MSX)
Galactic Plane Survey (Price et al. 2001). The MSX Band A includes
the unidentified infrared bands (UIBs) at 7.7 $\mu$m and 8.6 $\mu$m
with an angular resolution of about 18$^\prime$$^\prime$. And
near-IR data were obtained from the Two Micron All Sky Survey (2MASS) Point Source
Catalog (Skrutskie et al. 2006).

The SCUBA Legacy Catalogues (Di Francesco et al. 2008) provide two
comprehensive sets of continuum maps (and catalogs), using data at
850 and 450 $\mu$m obtained with the Submillimetre Common User
Bolometer Array (SCUBA), with angular resolutions of
19$^\prime$$^\prime$ and 11$^\prime$$^\prime$ respectively. The data
was reduced with the ``matrix inversion'' method described by
Johnstone et al. (2000). Objects are named by their respective
J2000.0 position of the peak 850$\mu$m intensity. The catalogues also
provide for each object the respective maximum 850 $\mu$m intensity,
estimates of total 850 $\mu$m flux and size, and tentative
identifications from the SIMBAD Database.

On the May of 2014, we performed CO observations using the 13.7 m
millimeter telescope of Qinghai Station at the Purple Mountain
Observatory at Delingha, China. On-the-fly (OTF) observing mode was
applied, with nine-pixel array receiver separated by $\sim$
180$^\prime$$^\prime$. The receiver was operated in the sideband
separation of single sideband mode, which allows for simultaneous
observations of three CO isotope transitions, with $^{12}$CO (1-0)
in the upper sideband (USB) and $^{13}$CO (1-0) and C$^{18}$O (1-0)
in the lower sideband (LSB). The typical system temperature
$T_{sys}$ is between 132 K and 221 K during the observations. The angular resolution of the telescope is
about 58$^\prime$$^\prime$, with beam efficiency between 0.44 at
115GHz and 0.51 at 110 GHz. The mapping step is
30$^\prime$$^\prime$ and the pointing accuracy is
better than 5$^\prime$$^\prime$. A fast Fourier transform (FFT)
spectrometer was used as the back end with a total bandwidth of 1
GHz and 16384 channels. The velocity resolution is about 0.16 km
s$^{-1}$ at $\sim$ 110 GHz. The spectral data were reduced and
analyzed with CLASS and GREG software.

\section{Results and analysis}
Fig.1 displays the images of Sh2-163 at different wavelengthes. The
upper panel shows the radio continuum emission at 1420 MHz from the CGPS. An
arc of strong radio continuum emission overlaying an extended
diffuse emission can be noted. From north to south, the radio
emission decreases. One of the two ionizing stars found by Russeil
et al. (2007) is very close to the center, and the other is close to
the peak of the radio emission. The middle panel shows an overlay of the
emission at 1420 MHz (line contours) and the optical image
(grey-scale). We can see that the arc-like structure of radio continuum emission is
coincidental with the brightest optical region. The ionized region of Sh2-163 is
detected both by the optical and radio continuum observations.
However, inside the ionized region, the optical emission is weak
near the location of R.A. (2000) = 23h33m17.8s and Dec. (2000) =
60$^\circ$46$^\prime$07.7$^\prime$$^\prime$. Interstellar dust in a foreground cloud may be
responsible for the observed optical
absorption feature. We did find a CO cloud which is spatially
coincident with the area lacking optical emission, using the data
from FCRAO CO Survey of the Outer Galaxy (Heyer et al. 1998). The bottom panel shows a two-color
image of Sh2-163: DSS-R image (green) and MSX band A image (red). An
arc-like structure of enhanced mid-infrared emission is evident on
the north side, which is also just outside the enhanced optical
emission. Like those observed in many other HII regions, the
polycyclic aromatic hydrocarbons (PAHs) may be responsible for the
emission detected at 8.3 $\mu$m, suggesting the existence of a PDR
on the border of Sh2-173. The radio emission seems to penetrate into
the PDR, indicating the interactions between the HII region and the
mid-infrared shell.

\begin{table}
\begin{minipage}{30cm}
\caption{Observed parameters of the emissions shown in Figure 3.}
\begin{tabular}{lrrccc}
\hline
Cloud     & Emissions   &  $V_{lsr}$            & $T_{mb}$              & rms & FWHM \\
         &            &(km s$^{-1}$)      & (K)           & (k)   & (km s$^{-1}$)    \\
\hline
A  & $^{12}$CO (1-0)  & -42.7 & 20.7 & 0.52 & 3.0 \\
   & $^{13}$CO (1-0)  & -42.6 & 9.0 & 0.31  & 2.1\\
   & C$^{18}$O (1-0)  & -42.4 & 3.2 & 0.36  & 1.1\\
\hline
B  & $^{12}$CO (1-0)  & -45.0 & 23.3 & 0.66  & 2.9\\
   & $^{13}$CO (1-0)  & -45.0 & 7.8 & 0.33 & 1.8 \\
\hline
\end{tabular}
\label{tb:rotn}
\end{minipage}
\end{table}

\subsection{CO emissions}

\begin{figure}
\centerline{\psfig{file=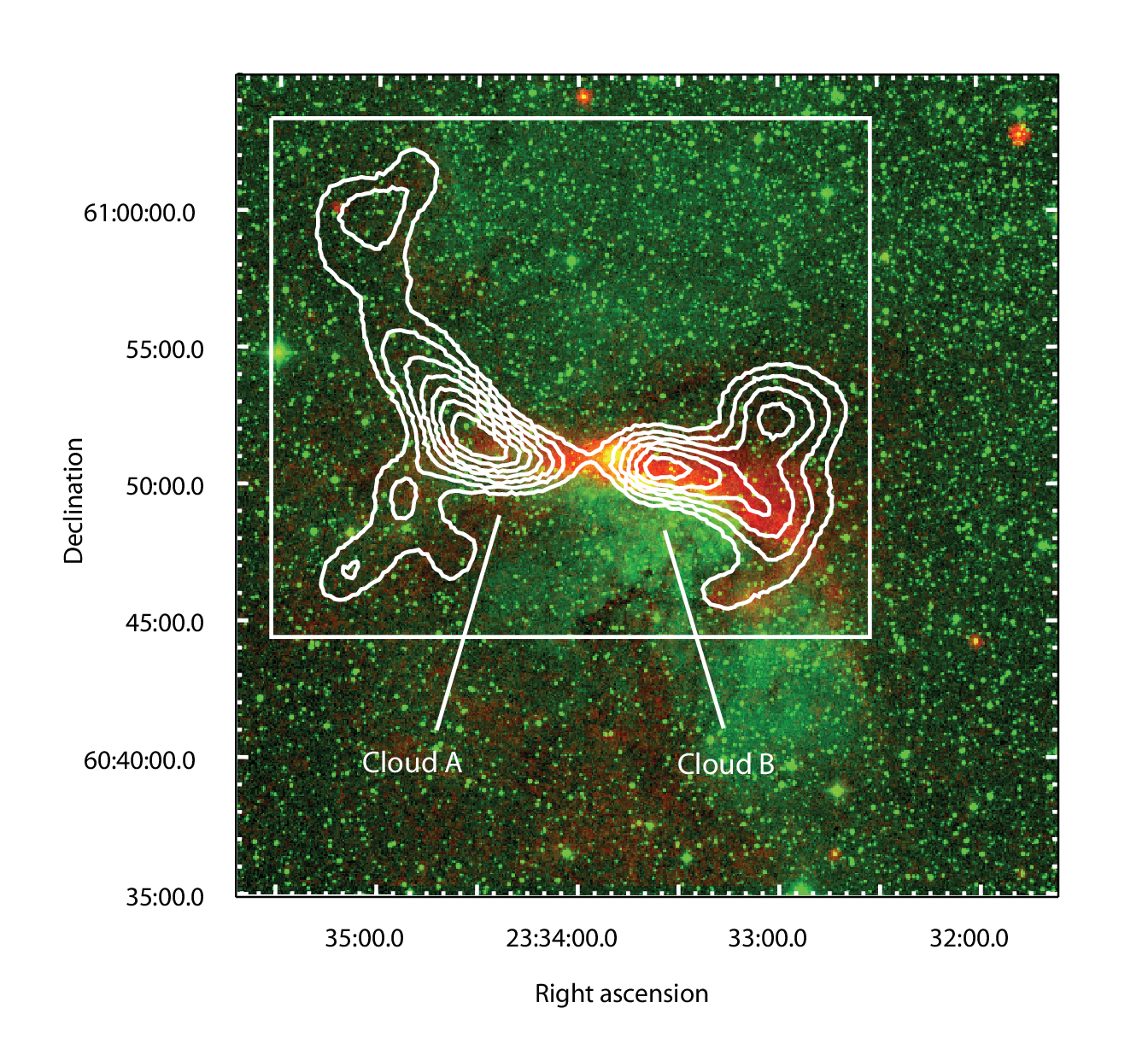,width=2.9in,height=2.5in}}
\caption{Integrated intensity contours of the FCRAO $^{12}$CO (1-0)
emission superimposed on the MSX band A (red) and DSS-R images (green). The contour levels
range from 10 to 40 K km s$^{-1}$ by 5 K km s$^{-1}$. The
integration range is from -39 to -52 km s$^{-1}$. The large
white box indicates the observing area by the 13.7 m telescope. }
\end{figure}

\begin{figure}
\centerline{\psfig{file=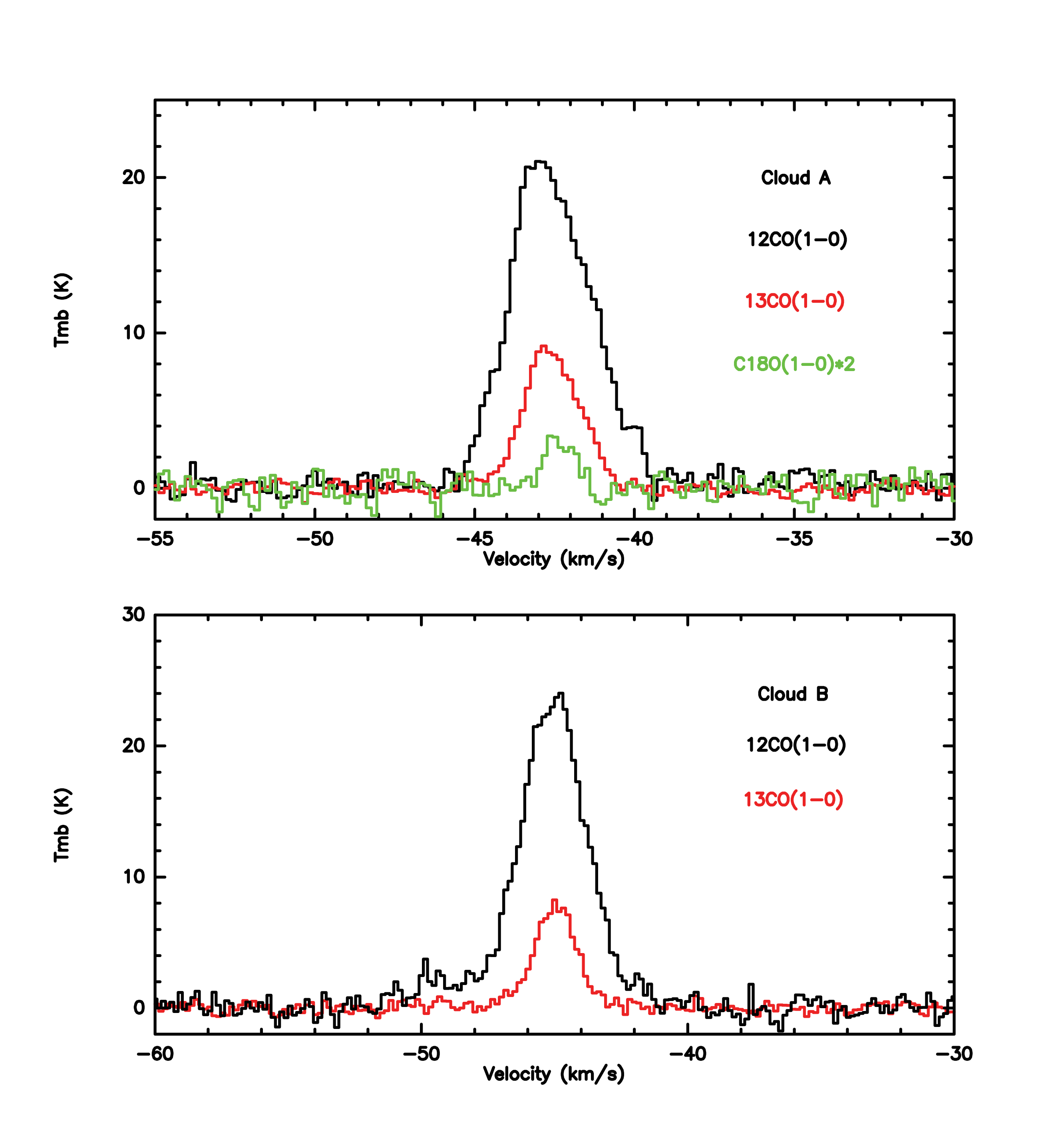,width=3in,height=3in}}
\caption{$^{12}$CO (black), $^{13}$CO (red), and C$^{18}$O (green) spectra at the peaks of molecular cloud A and B. The brightness
temperature of C$^{18}$O spectrum in cloud A is multiplied by 2.0. C$^{18}$O emission in cloud B was not detected.}
\end{figure}

\begin{figure}
\centerline{\psfig{file=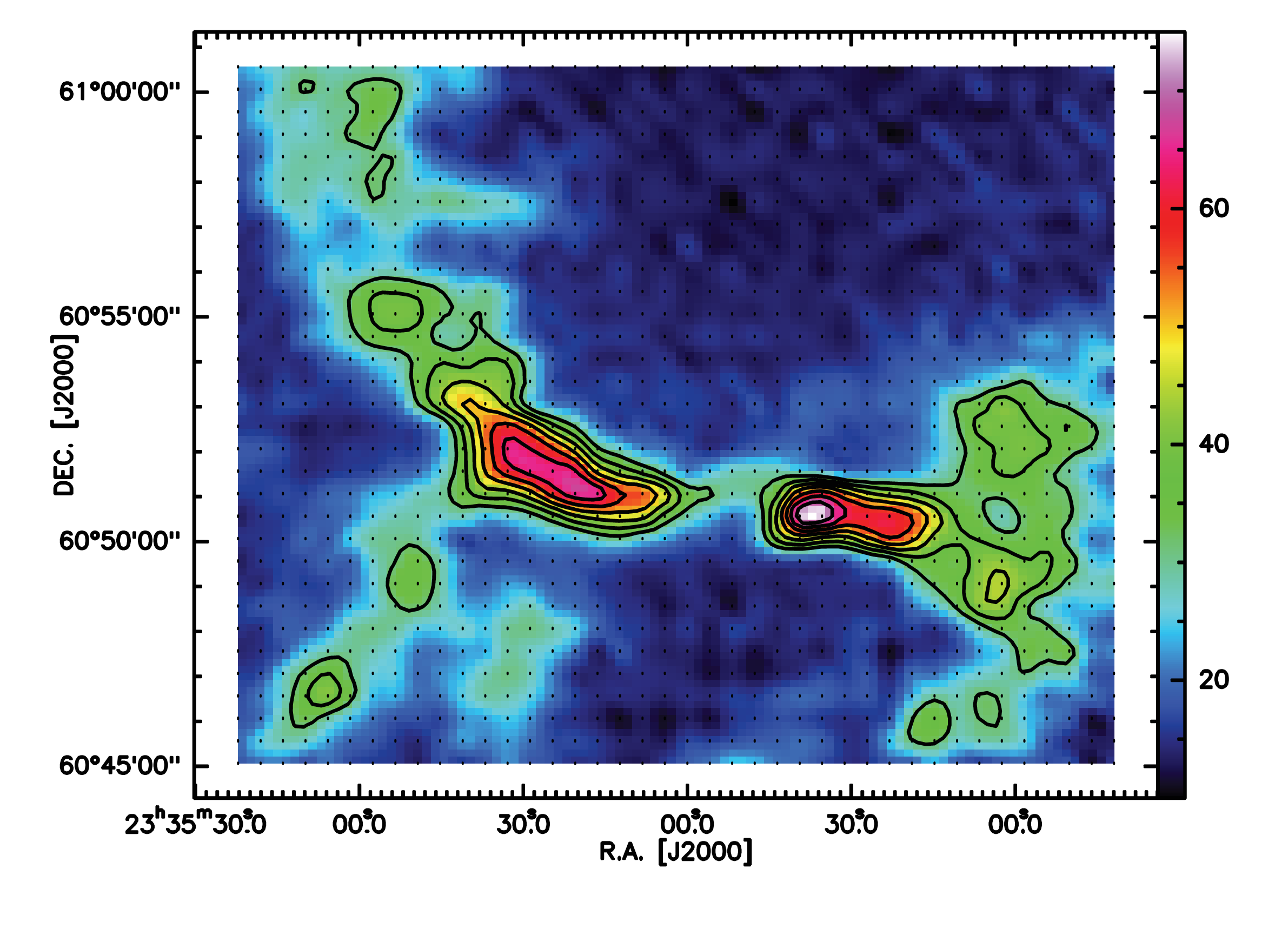,width=4in,height=2.5in}}
\centerline{\psfig{file=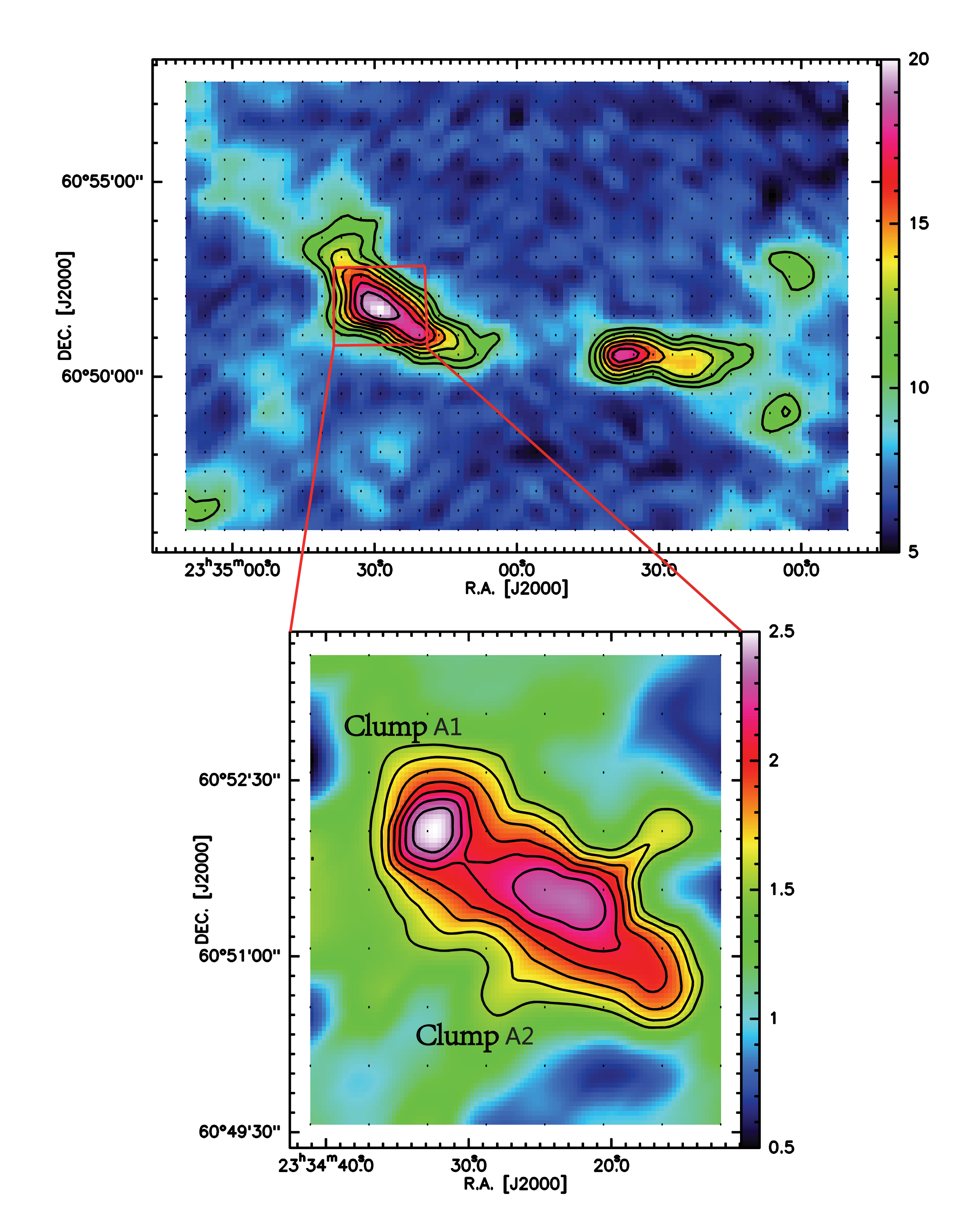,width=3in,height=4in}}
\caption{Top: Integrated intensity contours of $^{12}$CO from -50 km s$^{-1}$ to -38 km s$^{-1}$.
Middle: Integrated intensity contours of $^{13}$CO from -48 km s$^{-1}$ to -40 km s$^{-1}$.
Bottom: Integrated intensity contours of C$^{18}$O from -44 km s$^{-1}$ to -41 km s$^{-1}$.
Contour levels are 30$\%$, 40$\%$...90$\%$ of each peak emissions.}
\end{figure}

We first inspected the molecular gas around Sh2-163 from the CGPS
data in the whole velocity range and found an interesting feature
around $V_{lsr}$ $\sim$ -45 km s$^{-1}$. Fig.2 displays the integrated
intensity of $^{12}$CO (1-0) between -39 and -52 km s$^{-1}$ superimposed on
the MSX band A and DSS-R images. On the border
of Sh2-163, two molecular clouds were found with an arc-like morphology, indicating
they are compressed by the expanding HII region. The main peak of
each cloud is also consistent with the mid-infrared emission. We
further performed CO observations using the 13.7 m telescope to make
a detail study of these clouds. The resolution of 13.7 m telescope is
a little higher than that of the FCRAO 14m telescope (58$^{\prime\prime}$ vs.
100$^{\prime\prime}$). In addition, by studying the more optically thin lines of $^{13}$CO
and C$^{18}$O (if detected), we suppose to inspect the inner region of these clouds.

Fig.3 shows the CO isotope transitions at the peaks of molecular cloud A and B.
C$^{18}$O emission in cloud B was not detected. By fitting the C$^{13}$O lines with
gaussian functions, we obtained the peak velocities and FWHM (table 1). The derived V$_{lsr}$ of cloud A and B
is -42.6 km/s and -45.0 km/s, respectively. Fig.4 displays the integrated intensities of $^{12}$CO,
$^{13}$CO and C$^{18}$O lines. The morphology of two clouds is consistent with that detected by FCRAO.
Moreover, in cloud A we found two molecular clumps (noted by clump A1 and clump A2 in figure 4) by C$^{18}$O emissions. Detail study of the two clouds
will be discussed in section 3.3.
\begin{table}
\begin{minipage}{50cm}
\caption{Derived parameters of the two clouds.}
\begin{tabular}{clllcccc}
\hline
Cloud     & R.A.   & Dec.                    & $T_{ex}$              & $N(H_2)$  & Mass \\
         &  (J2000)      &  (J2000)   &(K)      & (cm$^{-2}$)               &    ($M_\odot$) \\
\hline
A  & 23h34m29s   & 60$^\circ$51$^\prime$40$^\prime$$^\prime$                    &  24.1                & 8.4 $\times$ 10$^{21}$          & 1341 \\
B  & 23h33m35s   & 60$^\circ$50$^\prime$34$^\prime$$^\prime$                   &  27.0                & 7.9 $\times$ 10$^{21}$            & 591  \\
\hline
\end{tabular}
\label{tb:rotn}
\end{minipage}
\end{table}

We now try to estimate the molecular column densities and hence
the masses of the two clouds from the $^{12}$CO and $^{13}$CO
observations. Under the assumptions of local thermodynamic equilibrium (LTE) and
$^{12}$CO to be optically thick, the excitation temperature (T$_{ex}$) can be obtained through each peak brightness temperature
of $^{12}$CO, via:

\begin{equation}
T_{ex} = \frac{5.653}{ln(\frac{5.653}{T_{mb} + 0.837} + 1)}
\end{equation}
The excitation temperature of clouds A and B is 24.1 K and 27.0 K, respectively. The total column densities of
$^{13}$CO can be obtained assuming that the $^{13}$CO emission is optically thin given by (Rohlfs $\&$ Wilson
2004):

\begin{equation}
N(^{13}CO) = 3.0 \times 10^{14}
\frac{T_{ex}}{1 - exp(-5.3/T_{ex})} \int \tau dv
\end{equation}
and
\begin{equation}
\int \tau dv = \frac{1}{J(T_{ex}) - J(T_{bg})} \int T_{mb} dv
\end{equation}
where T$_{bg}$ is the temperature of the background radiation (2.73
K). The column densities of H$_2$ could be
obtained by adopting typical abundance ratios [H$_2$]/[$^{12}$CO] = 10$^4$,
and [$^{12}$CO]/[$^{13}$CO] $\sim$ [$^{12}$C]/[$^{13}$C]. We adopted the Galactocentric
distance-dependent [$^{12}$C]/[$^{13}$C] ratio from Wilson $\&$ Rood (1994):
\begin{equation}
\frac{^{12}C}{^{13}C} = 7.5 \times R_{GC}[kpc] + 7.6.
\end{equation}
Using $M$ = $\mu$ \itshape m$_H$ $D^2$ \upshape $\Omega$
$N (H_2)$ we obtain the masses for cloud A and B, where $N (H_2)$ is the H$_2$ column
density calculated through the above equations, $D$ = 2.8 kpc
is the distance, $\Omega$ is the area of the clouds (within 50$\%$ of
each peak emission), and \itshape m$_{H}$ \upshape is the hydrogen
atom mass. We adopt a mean molecular weight per H$_2$ molecule of
$\mu$ = 2.72 to include helium. The derived parameters are listed in table 2.

\subsection{2MASS YSO candidates}
Both observations and theories indicate expanding HII regions may trigger next generation of stars (e.g. Osterbrock 1989;
Cichowolski et al. 2009; Miao et al. 2009; Panwar et al. 2014 ).
To look for evidence of triggered star formation, we searched for young stars in this region. We have adopted the criteria
developed by Kerton et al. (2008) and converted this criteria to the distance of Sh2-163 and the
visual absorption in this direction. From the classical relations, $N(H+H_2)$/$E(B-V)$ = 5.8 $\times$ 10$^{21}$
particles cm$^{-2}$ (Bohlin et al. 1978) and $A_V$ = 3.1$E(B-V)$, we obtain $A_V$ $\sim$ 5.34 $\times$ 10$^{-22}$
$N(H_2)$ $\sim$ 4.3 mag.
According to different photometric
qualities of 2MASS, we divided
their YSO candidates into four groups (P$_{1}$, P$_{1+}$, P$_{2}$,
P$_{3}$).  P$_{1}$ sources should have the valid photometry
in all three bands (i.e. ph$_{-}$qual values = A, B, C or D). For
sources in group P$_1$, the color
criteria are (J-H)$>$0.872 and (J-H)-1.7(H-K)+0.083$<$0. It selects
stars lying below the reddening vector associated with an O6V star.
P$_{1+}$ sources also have the valid photometry in all three bands.
The color criteria of group P$_{1+}$ are 1.172$<$(J-H)$<$1.472, (J-H)-1.7(H-K)+0.083$>$0,
(J-H)-1.7(H-K)-0.3797$<$0, and K$>$14.5. It selects YSO candidates lying in
the overlapping region of T Tauri and main sequence stars. Sources in
group P$_{2}$ have not been detected in the J band. Thus the actual
positions of P$_{2}$ sources in the (J-H) axe should be towards
higher values. The color criteria of group P$_2$ are
(J-H)-1.7(H-K)+0.083$<$0 and (H-K)$>$0.918. Sources belonging to
P$_{3}$ group have J and H magnitudes that are lower limits so their
color (J-H) can not be considered. The P$_{3}$ color criteria is
(H-K)$>$0.918. Following such criteria, we searched for tracers of
stellar formation activity in the 2MASS catalogue. Fig.5 shows the
color-color diagram (CCD) of the selected YSO candidates. The two parallel lines
are reddening vectors, assuming the interstellar reddening law of
Rieke \& Lebofsky (1985) (A$_{J}$ / A$_{V}$ = 0.282; A$_{H}$ /
A$_{V}$ = 0. 175; A$_{K}$ / A$_{V}$ = 0.112). Fig.6 shows the locations of the YSO candidates.

\begin{figure}
\centerline{\psfig{file=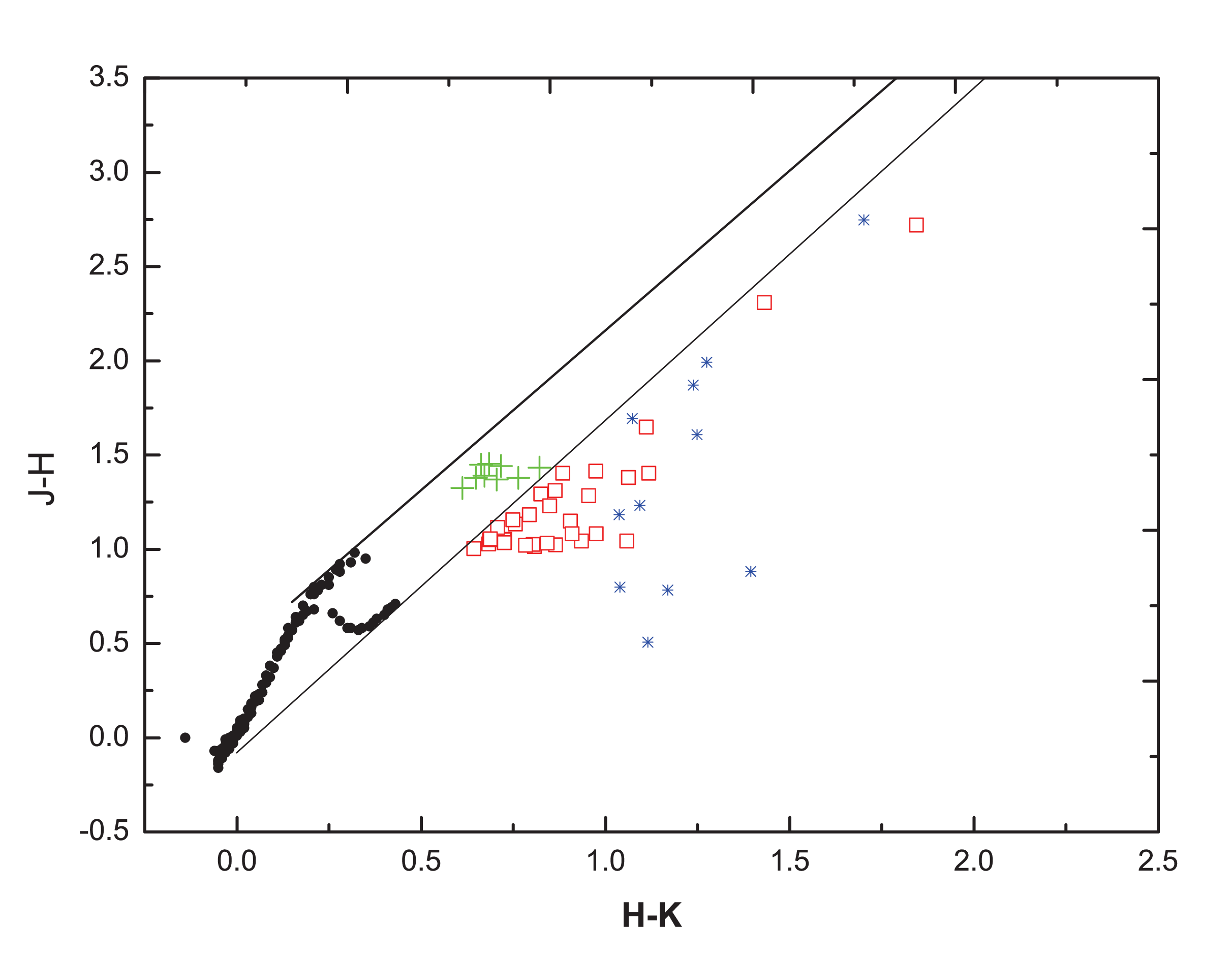,width=3in,height=2.5in}}
\caption{CC diagram of the 2MASS YSO candidates around
Sh2-163. P$_{1}$, P$_{1+}$, P$_{2}$ sources are indicated by red
boxes, green crosses, and blue asterisks, respectively. The locations of
main sequence and giant stars are derived from Koornneef (1983). The
two parallel solid lines are reddening vectors. }
\end{figure}

\begin{figure}
\centerline{\psfig{file=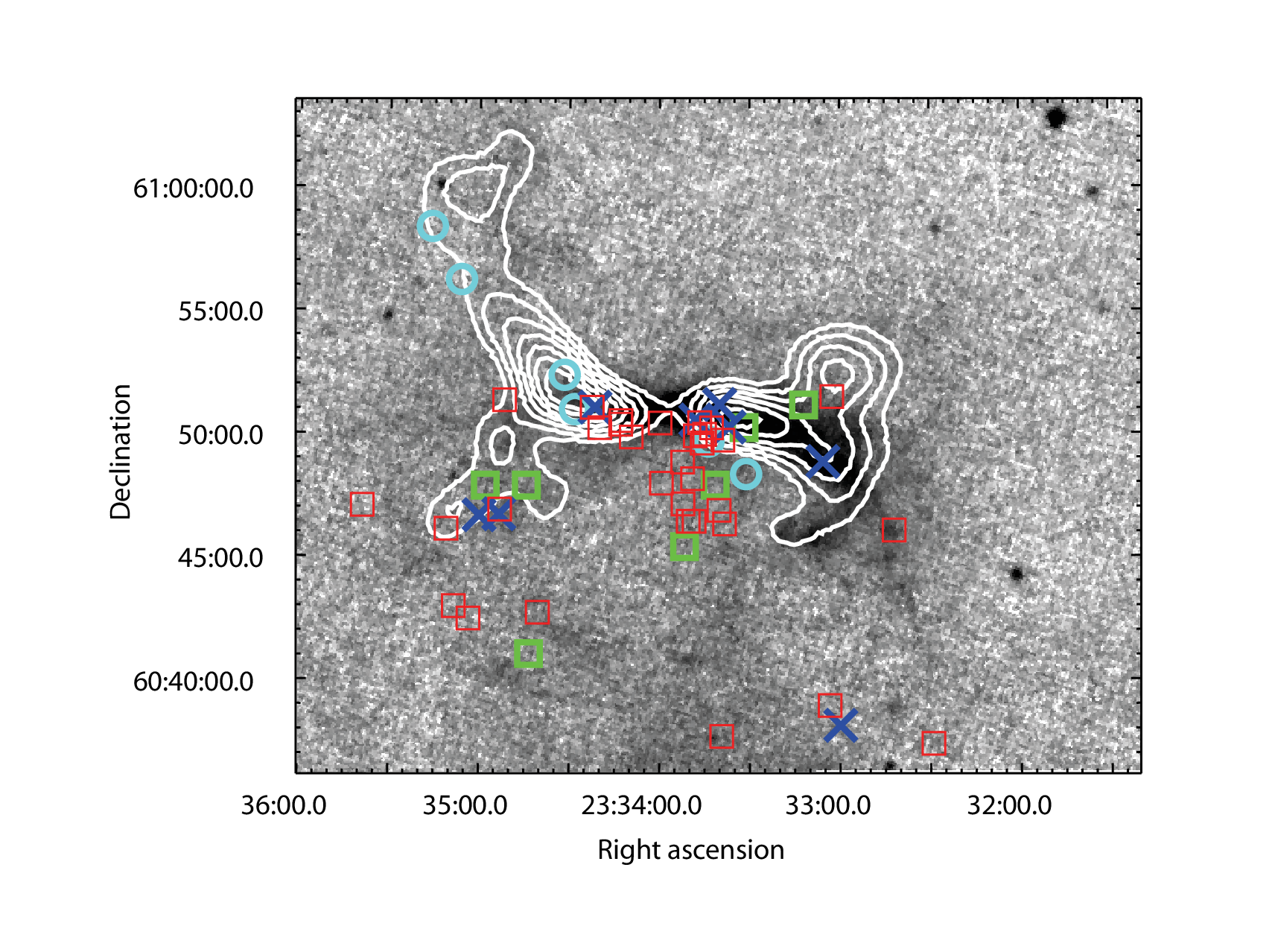,width=3in,height=2.5in}}
\caption{The YSO candidates found through 2MASS point catalogue. P$_{1}$, P$_{1+}$,
P$_{2}$, P$_{3}$ sources are indicated by red boxes, green boxes,
blue X points, and cyan circles respectively. The contour levels are
the same as those in Fig.2. }
\end{figure}

We now discuss the likelihood of triggering star formation in
Sh2-163. It can be noted the presence of a group of these sources
right upon the north side of Sh2-163. For the two molecular clouds
we discussed above, there is a very good relationship between CO emission,
infrared shell and YSO candidates. Almost no YSO candidates was found
outside the shell, even though there are still strong CO emissions.
Such phenomenon suggest triggered star formation may be taking
place in this region by the expansion of Sh2-163. The morphology of
these clouds detected by CO emissions could not be purely explained by the
``collect and collapse'' model, as the CO emission distribution suggests the presence of pre-existing molecular clouds in the border of the HII region.
The DSS-R image of Figure 1
neither displays cometary morphology. We thus discard the so-called
``RDI'' and ``C\&C'' processes. Another possibility is that the
shocked expanding layer, prior to the beginning of the instability,
collides with these pre-existing molecular clouds (A and B). Star formation would
take place at the interface between the layer and the cloud clumps. Sh2-163 is not the only object; HII regions like Sh2-
235 (Kirsanova et al. 2008), Sh2-217 and Sh2-219 (Deharveng et al. 2003) also show such physical
processes of sequential star formation.

\begin{table*}
\begin{minipage}{10cm}
 \caption{\label{tab:test}Outflow parameters.}
 \begin{tabular}{clllcccc}
  \hline
Shift & Integrated range & N (CO)  & M$_{out}$ & P$_{out}$ & E$_{out}$
\\
& (km s$^{-1}$)& ($\times$ 10$^{16}$ cm$^{-2}$)& ($M_\odot$) &  ($M_\odot$ km s$^{-1}$) &(M$_\odot$ [km s$^{-1}$]$^2$)\\
  \hline
 red &(-43.5, -39)& 1.9 & 7.6&46&137\\
 blue&(-46.5, -51)& 2.7& 14.2&85&255\\
  \hline
 \end{tabular}
  \label{tb:rotn}
\end{minipage}
\end{table*}

\subsection{Star formation in cloud A and B}
\begin{figure}
\centerline{\psfig{file=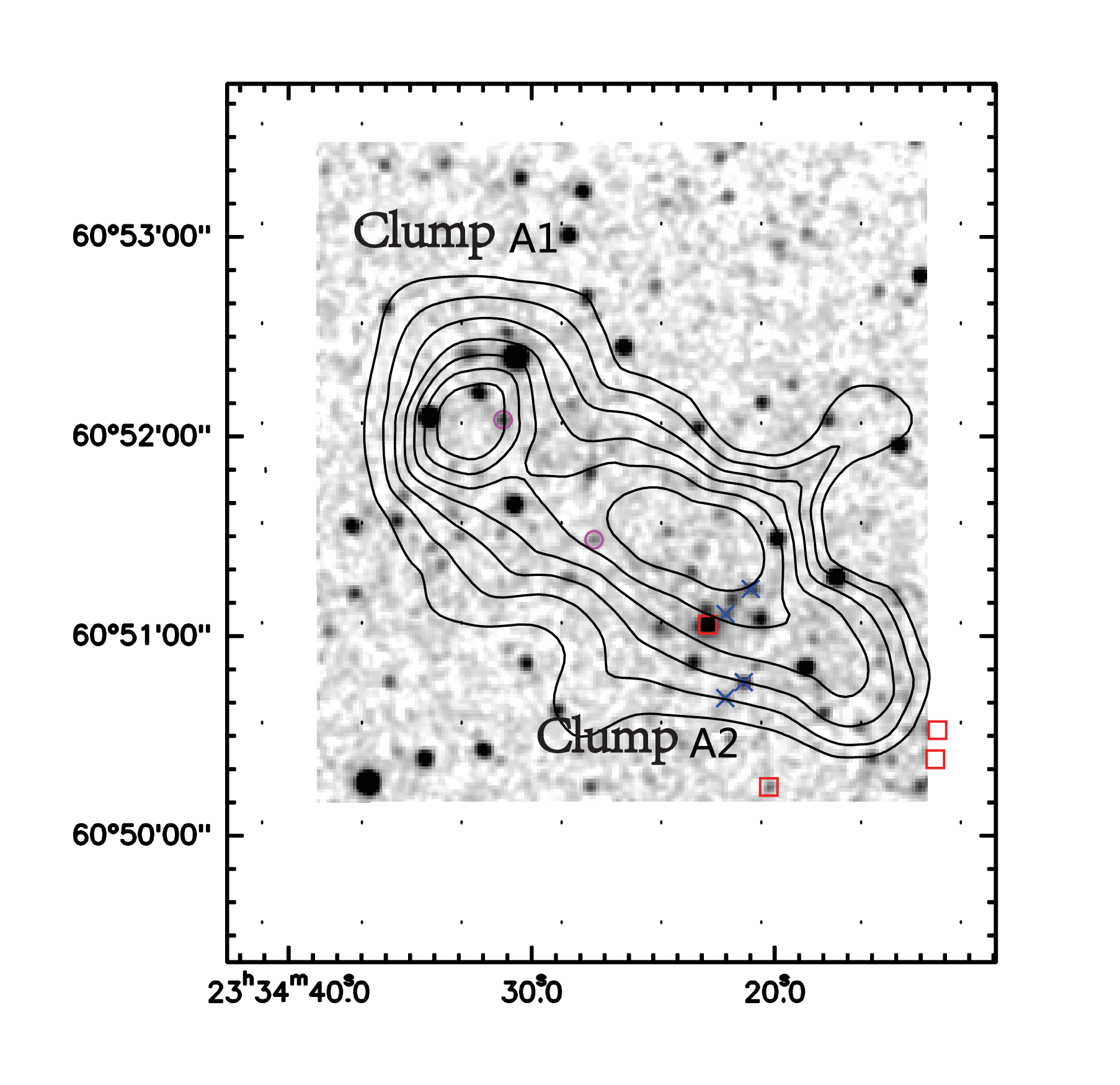,width=3in,height=2.5in}}
\centerline{\psfig{file=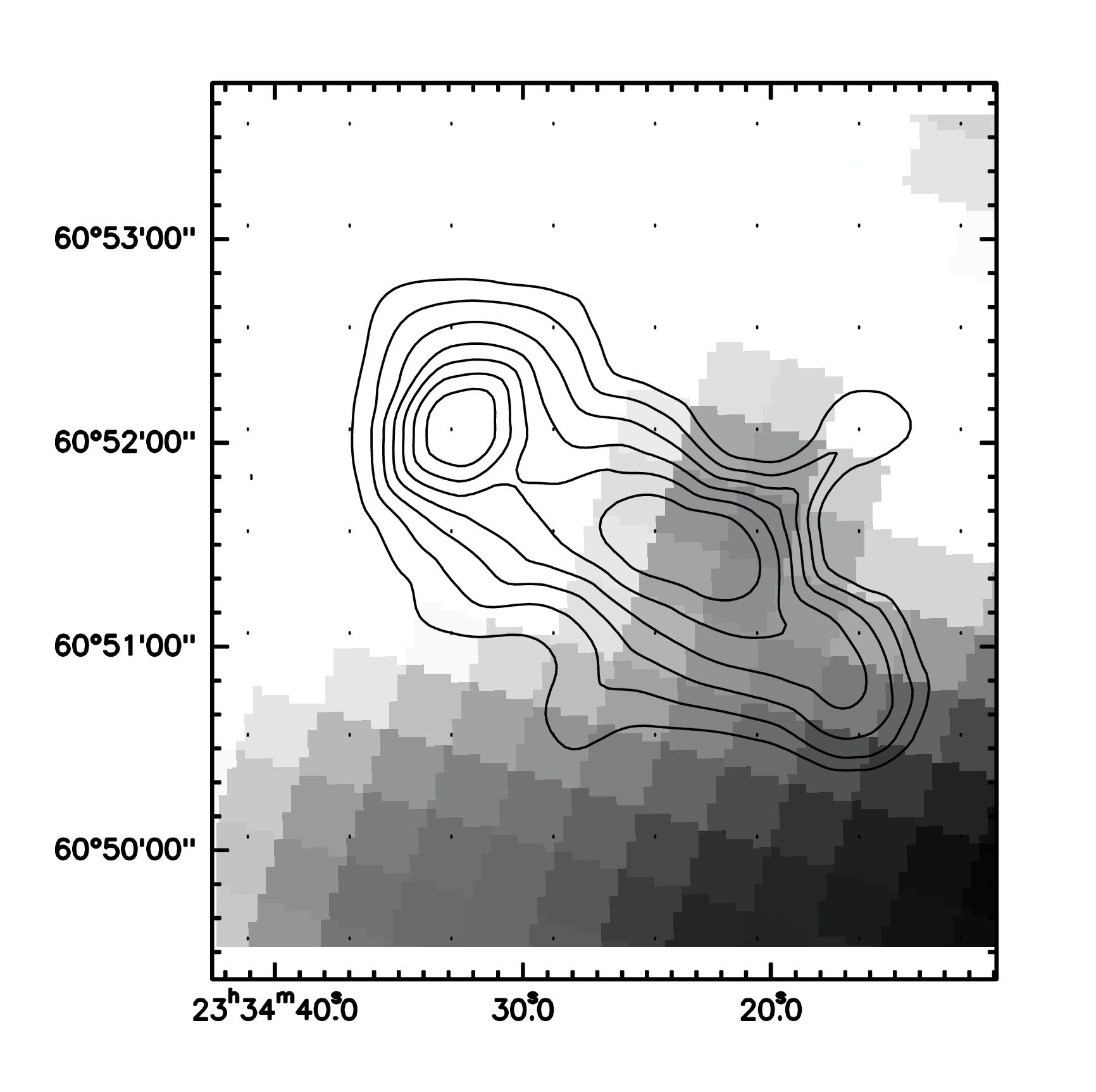,width=3in,height=2.5in}}
\caption{ Top: NIR $K$-band image (gray-scale) with the C$^{18}$O intensity (contours)
overlaid. Contour levels are the same as those in Fig.4. Symbols of YSO candidates are
the same as those in Fig.6. Bottom: 1420 MHz image (grey-scale) with the C$^{18}$O intensity (contours)
overlaid.  }
\end{figure}

\begin{figure}
\centerline{\psfig{file=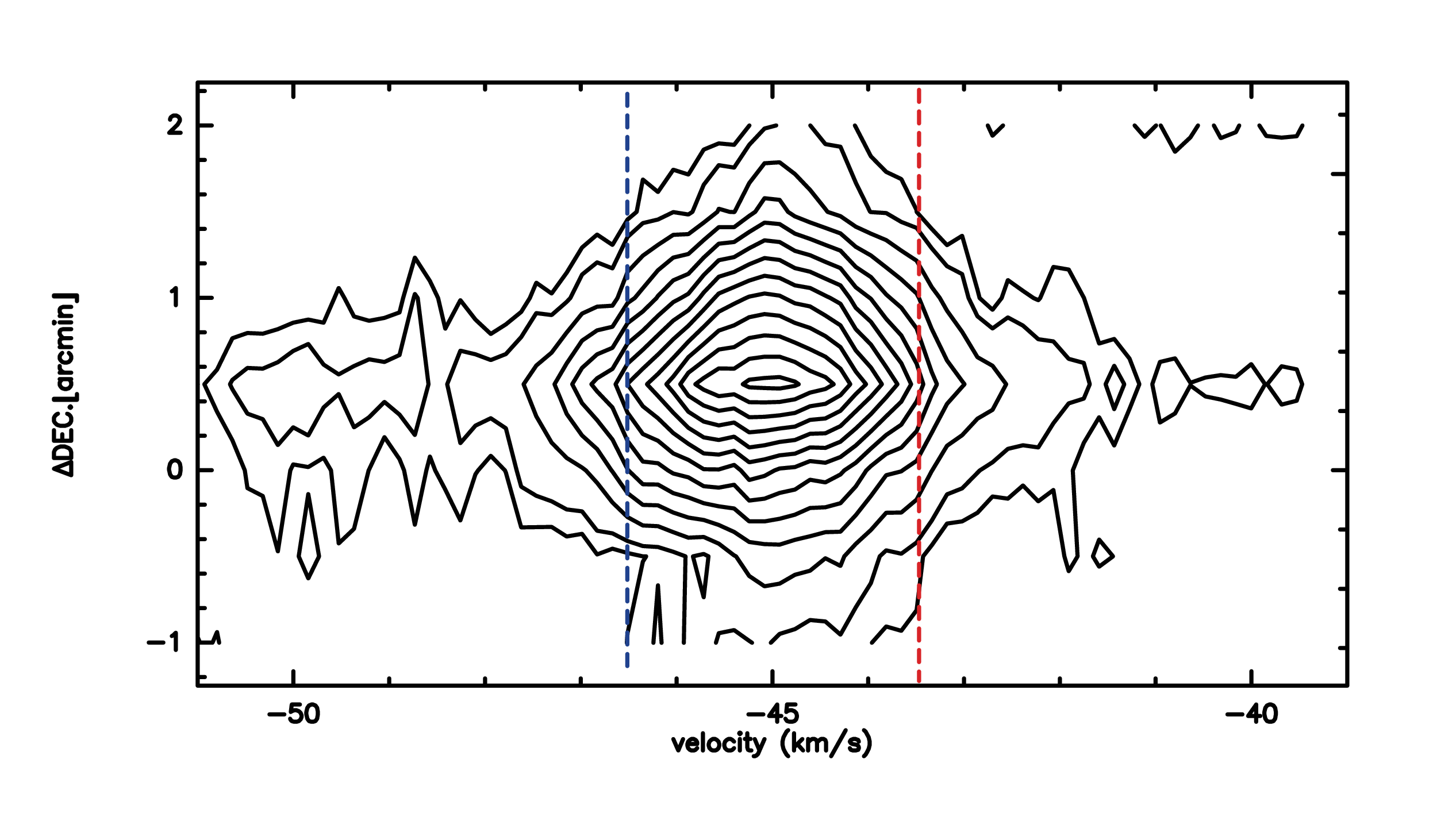,width=3in,height=2.5in}}
\caption{PV diagram of cloud B cut from north to south. Contours are 1 to 25 K by step of 2 K. }
\end{figure}

\begin{figure}
\centerline{\psfig{file=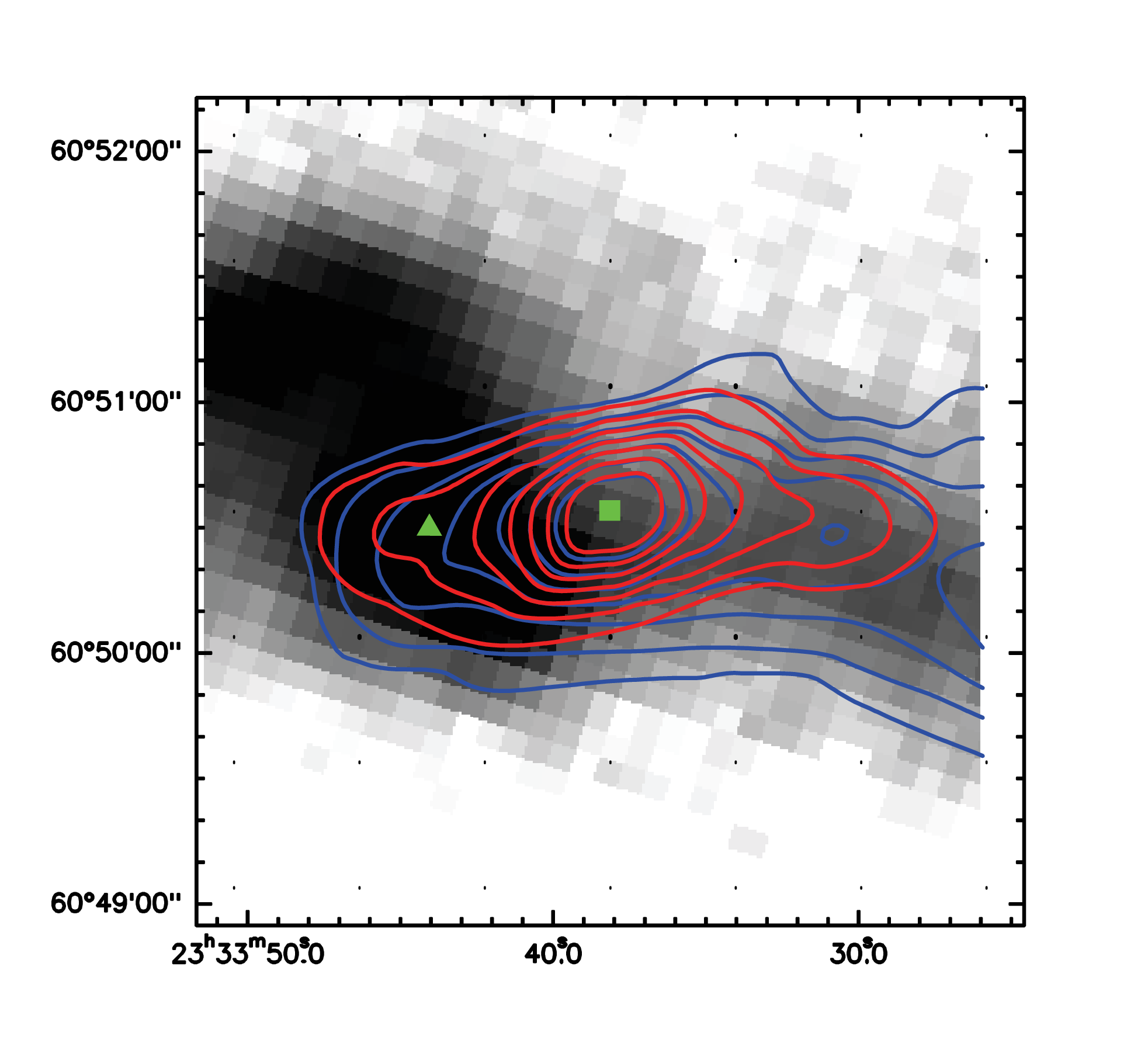,width=3in,height=2.5in}}
\centerline{\psfig{file=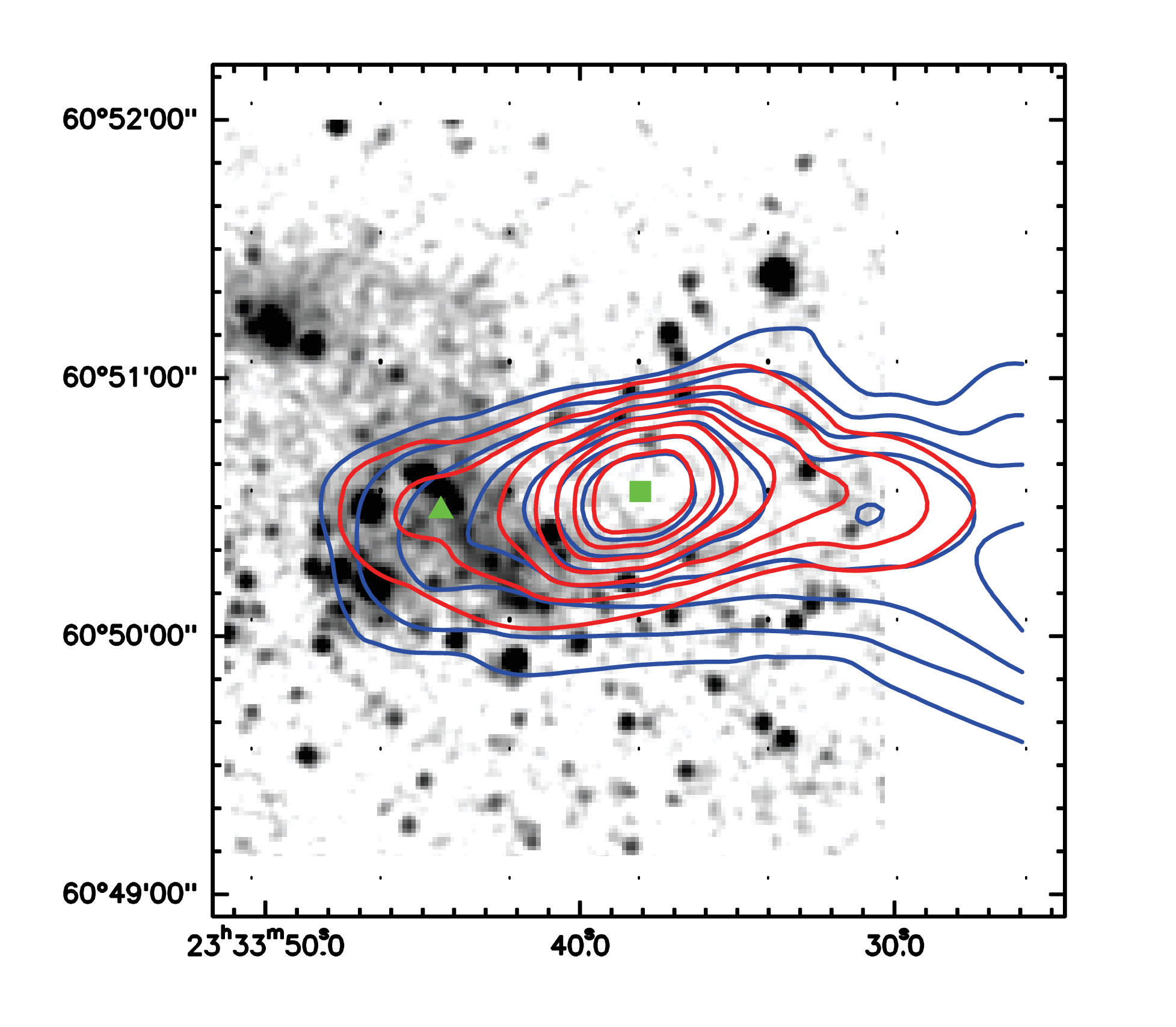,width=3in,height=2.5in}}
\caption{The CO outflow (contours) with MSX band A (upper) and 2MASS $K$-band image (bottom). The outflow lobes
are contoured from 30$\%$ of the peak emission in steps of 10$\%$. The green triangles mark the location of IRAS 23314+6033.
The green boxes mark the center of cloud B.}
\end{figure}
We now make a detail study of star formation in cloud A and B. As
mentioned above, two molecular clumps were found in cloud A by C$^{18}$O
observations. Fig.7 shows the integrated intensity of C$^{18}$O superimposed
on the near-infrared (NIR) $K$-band image of 2MASS. We can note that
there are six 2MASS YSO candidates projected onto clump A2, while only one candidate
in clump A1. Fig.7 also shows the 1420 MHz image (grey-scale) with the C$^{18}$O intensity
(contours) overlaid. It seems star formation in A2 is much more active than in A1. One good
reason for this phenomenon is that the shock passed A2 first (triggered star formation
inside) and then to A1.
However, further observations should be carried out to study our speculation.

We found outflow activities in cloud B. Fig. 8 shows the $^{12}$CO position-velocity (PV) diagram of cloud B
cut from north to south. The wing emission is obvious in the PV diagram. Typical outflows appear
as spatially confined wings beyond the emission from the cloud. According to the PV diagram,
we selected the integrated range of wings and determined the outflow intensities of red
and blue lobes(figure 9).
Using the method described in section 3.1, we obtain the masses for the red and
blue lobes of the outflow. We estimate the momentum and energy
of the red and blue lobes using
\begin{equation}
P_{out} = M_{out} V
\end{equation}
and
\begin{equation}
E_{out} =\frac{1}{2} M_{out} V^2
\end{equation}
where $V$ is a characteristic velocity estimated as the difference
between the maximum velocity of CO emission in the red and blue
wings respectively, and the molecular ambient velocity ($V_{lsr}$).
The derived parameters are shown in table 3.

Near the core of cloud B we found that the IRAS point source
23314+6033 satisfied the protostellar object colors described by Junkes
et al. (1992): S$_{100}$ $\ge$ 20 Jy, 1.2 $\le$
$\frac{S_{100}}{S_{60}}$ $\le$ 6.0, $\frac{S_{60}}{S_{25}}$ $\ge$ 1
and Q$_{60}$ + Q$_{100}$ $\ge$ 4, where S$_{\lambda}$ and
Q$_{\lambda}$ are the flux density and the quality of the IRAS flux
in each of the observed band respectively. The total infrared
luminosity of IRAS 23314+6033 can be calculated by the method of
Casoli et al.\,(1986):
\begin{equation}
\begin{split}
&L_{IR}=(20.65\times S_{12}+7.35\times S_{25}+4.57\times
S_{60}+1.76\times S_{100})\\
&\times D^2\times0.30
\end{split}
\end{equation}
where D is the distance from the solar system in kpc.
According to Smith et al. (2002), the luminosity of 17380
L$_{\odot}$ corresponds to a B1.5 star.
Zhang et al. (2005) made a $^{12}$CO(2-1) observation of IRAS 23314+6033 (1$^\prime$ $\times$ 1$^\prime$ in step of 29$^\prime$$^\prime$), using
the 12 m telescope of the National Radio Astronomy Observatory (NRAO) at Kitt Peak. They discovered an unresolved outflow driven by
the IRAS point source. IRAS 23314+6033 is associated with the Red MSX Source (RMS)
G113.6041-00.6161 (Lumsden et al. 2002), which is about 45$^\prime$$^\prime$ away from the peak of cloud B (the upper panel in figure 9).
Considering the spacial resolution of MSX,
we do not regard IRAS 23314+6033 is on the center of cloud B.
Our larger area observations indicate the outflow discovered by Zhang et al. (2005) seems
to be just a part of the main outflow from cloud B.
Young star(s) deeply embedded on the center of cloud B drive(s)
the main outflow. The NIR H$_2$ emission is a good tracer of shocks in molecular outflows
in low-mass star-forming regions.
We can see extended emission in the $K$ band in figure 9, which may be
partly due to excited H$_2$ emission at 2.12 $\mu$m.

\begin{figure}
\centerline{\psfig{file=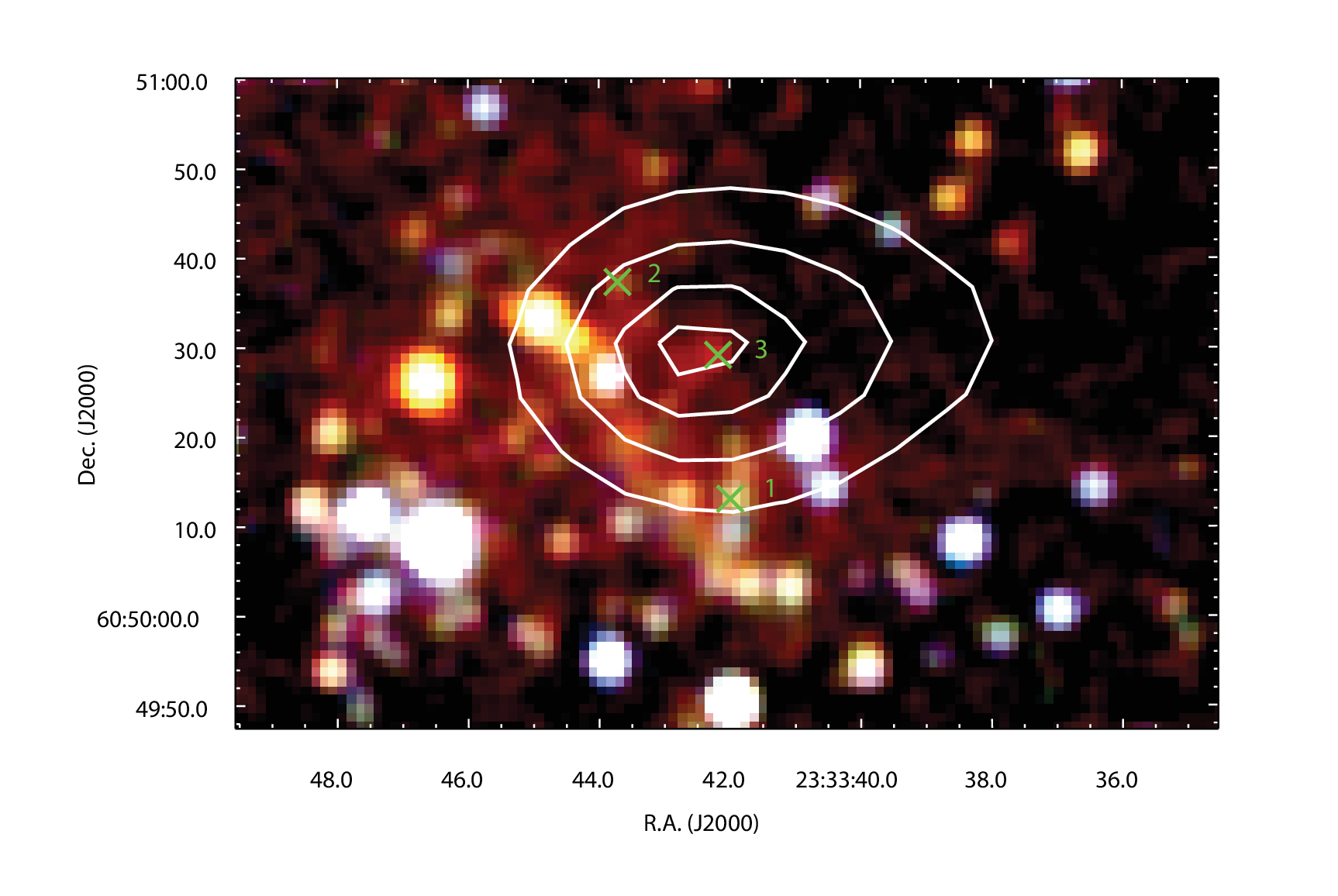,width=3in,height=2.5in}}
\caption{RGB composite of 2MASS J-band (blue), H-band (green) and
K-band (red) images of the 850 $\mu$m source. The contours
correspond to 850 $\mu$m brightness levels of 0.9, 1.2, 1.5, 1.8 Jy
beam$^{-1}$, 1$\sigma$ = 0.04 Jy beam$^{-1}$.  }
\end{figure}

\begin{figure}
\centerline{\psfig{file=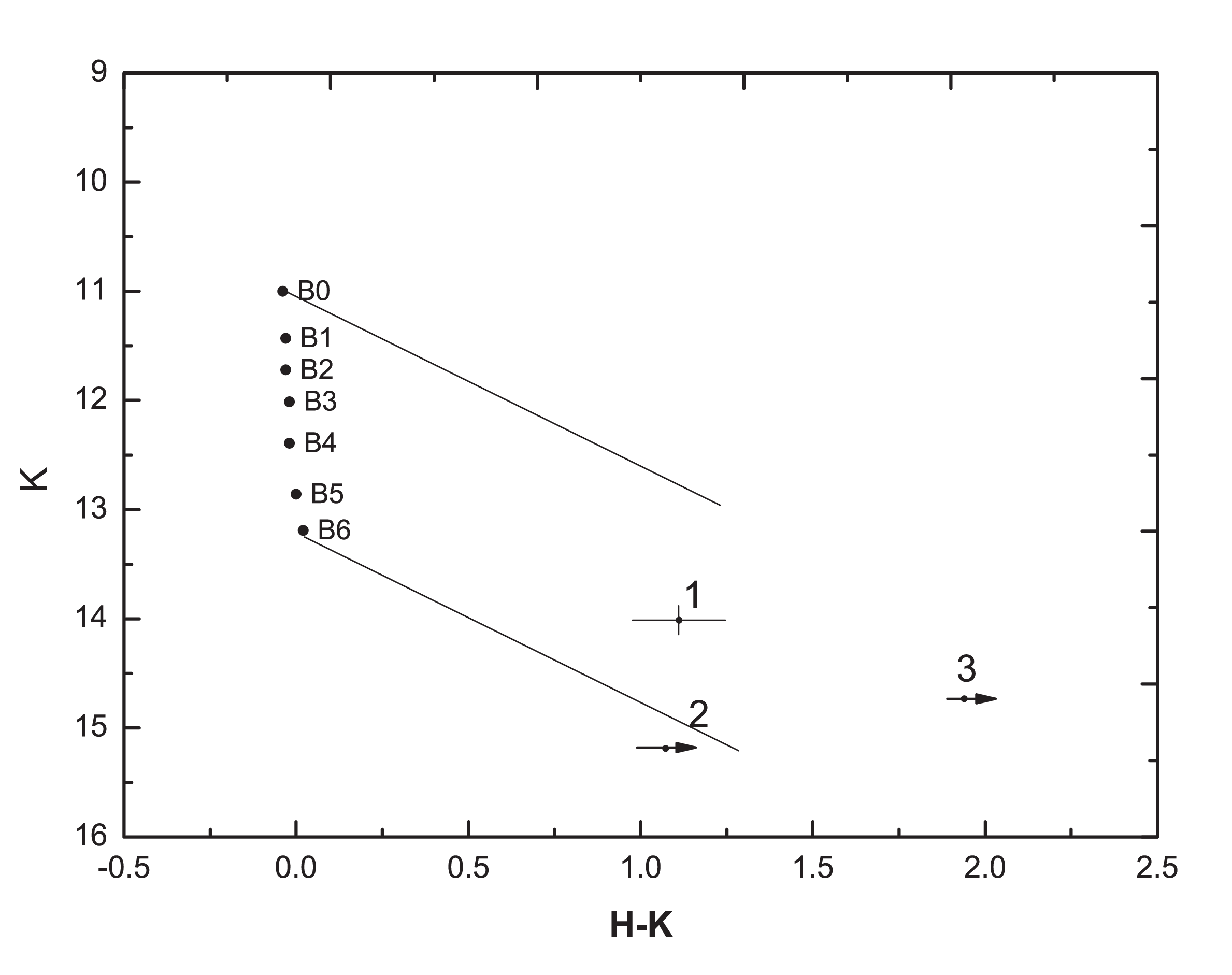,width=3in,height=2.5in}}
\caption{Color-magnitude diagram for the 2MASS YSO candidates whin
the sub-mm source. The position of the zero-age-main-sequence stars
with A$_v$ = 0 mag at a distance of 2.8 kpc are indicated by filled
diamonds. The parallel lines are A$_v$ = 20 reddening vectors.  Two
arrows indicate sources where the 2MASS catalog only lists an upper
brightness limit in the relevant band. }
\end{figure}

We also searched for sources in the SCUBA legacy catalogues and found a prominent
sub-mm source J233342.7+605030 associated with IRAS 23314+6033. The
maximum brightness (B$_{850}$) of J233342.7+605030 is 1.91 Jy
beam$^{-1}$, with a flux density (F$_{850}$) of 3.3 Jy (Di Francesco
et al. 2008). The deconvolved radius is 34.2 arcsec and corresponds
to a physical diameter of 0.54 pc. Using the 850 $\mu$m continuum
emission, we estimate the dust mass from the relationship of Tej et
al. (2006):
\begin{equation}
M_{dust}=1.88\times10^{-4}(\frac{1200}{\nu})^{3+\beta} S_{\nu}
(e^{0.048\nu/T_d}-1) D^2
\end{equation}
where S$_{\nu}$ is the flux density at the frequency $\nu$. We also
assume that the dust temperature T$_d$ is 20 K, $\beta$ the dust
emissivity index is 2.6 for the assumed dust temperature in this
region according to Hill et al. (2006), and D the distance to
Sh2-163 in kpc. We thus obtain a dust mass of M$_{dust}$ $\sim$ 9.1
M$_{\odot}$.

Fig.10  shows a composite image of the region surrounding the
850$\mu$m source, made from 2MASS J-band (blue), H-band (green) and
K-band (red) images. The contours, 850 $\mu$m brightness levels of
0.9, 1.2, 1.5, 1.8 Jy beam$^{-1}$, trace the very central part of
the sub-mm source. Three YSO candidates selected through the
criteria described above locate within the sub-mm emission, with star 3 in the center. Fig.11
shows these stars plotted in a 2MASS color-magnitude diagram (CMD). The
main-sequence stars are shown with some representative spectral
types at a distance of 2.8 kpc. The two parallel lines are A$_{v}$ =
20 reddening vectors for a B0V and B6V star using the interstellar
reddening law of Rieke \& Lebofsky (1985). Arrows indicate sources
where the 2MASS catalog only lists an upper brightness limit in the
relevant band. The source labeled 3 in Fig.10 and Fig.11 is probably a
highly embedded YSO with spectral type between B0V and B3V,
consistent with the IR luminosity analysis above. Thus, we suggest
that the most likely massive protostar related to IRAS 23314+6033
would be star 3.

\section{summary}
Based on our CO emission observations of 13.7-m PMO telescope,
together with other archival data including CGPS,
2MASS, MSX, and SCUBA, we made a multi-wavelength study of Sh2-163.
The main results can be summarized as follows:

1.Radio continuum emissions, optical observations and mid-infrared images of Sh2-163 indicate the strong
interactions between the HII region and the surrounding ISM.

2.Two molecular clouds were discovered on the border of
PDR. The morphology of these two clouds also suggests they are compressed by
the expansion of Sh2-163. In cloud A, we found two molecular clumps.
Our study indicates star formation in clump A2 seems to be more active than in clump A1.
In cloud B, we found outflow activities driven by young star(s) still deeply embedded.

3.Using 2MASS photometry, we searched for
embedded YSO candidates in this region. The
very good relations between CO emission, infrared shell and YSOs
suggest that it is probably a triggered star formation region by the
expansion of Sh2-163. We discard the so-called ``RDI'' and ``C\&C'' processes taking place in Sh2-163, and propose another possibility that the
shocked expanding layer, prior to the beginning of the instability,
collides with these pre-existing molecular clouds (A and B). Star formation would
take place at the interface between the layer and the cloud clumps.

4.We found a prominent sub-mm source at the location of
IRAS 23314+6033. Three YSOs candidates were
found imbedded. On the peak of the sub-mm emission locates an
inter-mediate massive young star consistent with the IR luminosity
analysis. We thus regard having found the most likely massive
protostar related to IRAS 23314+6033.

\section*{ACKNOWLEDGEMENTS}
We thank the anonymous referee for the constructive suggestions and
we are also grateful to the staff at the Qinghai Station of Purple Mountain Observatory (PMO) for their
observations. The CGPS is a Canadian project with international partners
and is supported by grants from NSERC. Data from the CGPS are publicly
available through the facilities of the Canadian Astronomy Data Center
(http://cadc.hia.nrc.ca) operated by the Herzberg Institute of Astrophysics, NRC.

\label{lastpage}
\end{document}